\documentclass[twocolumn,showpacs,amsmath,amssymb,pra]{revtex4}

\usepackage{graphicx}
\usepackage{dcolumn}
\usepackage{bm}
\usepackage{subfigure}

\begin{document}


\title {
Towards an {\it ab initio} theory for the temperature dependence of electric-field gradient in solids:
application to hexagonal lattices of Zn and Cd
}

\author{A. V. Nikolaev}
\affiliation{Skobeltsyn Institute of Nuclear Physics Lomonosov Moscow State University, 119991 Moscow, Russia}
\affiliation{School of Electronics, Photonics and Molecular Physics, Moscow Institute of Physics and Technology, 141700, Dolgoprudny, Moscow region, Russia}
\affiliation{National Research Nuclear University MEPhI, 115409, Kashirskoe shosse 31, Moscow, Russia}

\author{N. M. Chtchelkatchev}
\affiliation{Vereshchagin Institute for High Pressure Physics, Russian Academy of Sciences, 108840 Troitsk, Moscow, Russia}
\affiliation{Ural Federal University, 620002, Ekaterinburg, Russia}
\affiliation{Moscow Institute of Physics and Technology, 141700, Dolgoprudny, Moscow Region, Russia}

\author{D. A. Salamatin}
\affiliation{Vereshchagin Institute for High Pressure Physics, RAS, 142190 Troitsk, Moscow, Russia}
\affiliation{Dzelepov Laboratory of Nuclear Problems, Joint Institute for Nuclear Research, 141980 Dubna, Russia}

\author{A. V. Tsvyashchenko}
\affiliation{Vereshchagin Institute for High Pressure Physics, RAS, 142190 Troitsk, Moscow, Russia}



\begin{abstract}
Based on {\it ab initio} band structure calculations
we formulate a general theoretical method for description of the temperature dependence of electric field gradient in solids.
The method employs a procedure of averaging multipole electron density component ($l \neq 0$) inside a sphere
vibrating with the nucleus at its center. As a result of averaging each Fourier component ($K \neq 0$) on the sphere is effectively reduced
by the square root of the Debye-Waller factor [$\exp(-W)$].
The electric field gradient related to a sum of $K-$components most frequently decreases with temperature ($T$), but under certain
conditions because of the interplay between terms of opposite signs it can also increase with $T$.
The method is applied to calculations of the temperature evolution of the electric field gradients of pristine zinc and cadmium
crystallized in the hexagonal lattice.
For calculations within our model of crucial importance is the temperature dependence of mean-square displacements which
can be taken from experiment or obtained from the phonon modes in the harmonic approximation.
For the case of Zn we have used data obtained from single crystal
x-ray diffraction. In addition, for Zn and Cd we have calculated mean-square displacements with the density functional perturbation treatment of
the Quantum Espresso package.
With the experimental data for displacements in Zn our calculations reproduce
the temperature dependence of the electric field gradient very accurately.
Within the harmonic approximation of the Quantum Espresso package the decrease of electric field gradients in Zn and Cd with temperature is overestimated.
Our calculations indicate that the anharmonic effects are of considerable importance in the temperature dependence of electric field gradients.
\end{abstract}


\maketitle

\section{Introduction}
\label{sec:int}

The electric field gradient (EFG) is a very sensitive characteristic of electron structure \cite{Kauf,Das,Haa0,Nuc}.
It is directly measured by the family of methods experiencing quadrupole hyperfine interactions,
such as nuclear quadrupole resonance (NQR), M\"{o}ssbauer spectroscopy (MS) and perturbed angular correlation (PAC) spectroscopy
\cite{Nuc,PAC,PAC17}.
Nuclear probes in these techniques are exposed to the
local electronic and molecular structure via the electric
interaction between the nuclear quadrupole moment and the
surrounding electronic charge distribution
providing a spectroscopic fingerprint of the electron environment.

The methods are utilized in a wide range of applications.
The time differential perturbed $\gamma-\gamma$ angular correlation (TDPAC) spectroscopy \cite{PAC,PAC17} for example
can be used in biochemistry characterizing interactions between metal ions and proteins \cite{PACbio},
point defects in metals and recently in semiconductors \cite{PACsemi}, surface and interface
properties, detecting charge density wave formations and structural phase transitions  \cite{Tsv}
in various materials.
Although the TDPAC spectroscopy has been known for more than 40 years, it still
has a rich potential for solid state physics and novel materials \cite{PAC17,PACbio,PACsemi}.

On the other hand EFGs measured by these techniques, can be confronted with theoretical values obtained
from {\it ab initio} calculations \cite{Blaha1,Blaha2,Petr}
which should give a thorough picture of microscopic properties of the investigated material.
Recently, such a comparison has lead to an improved value of the nuclear quadrupole moment of cadmium \cite{Err,Haa}.
The problem is however that in many cases there is a strong temperature ($T$) dependence of EFGs \cite{Kauf,Das,Haa0,Web},
discussed in detail in Sec.\ \ref{theories} below,
which is often not taken into account.
In contrast to {\it ab initio} calculations of EFGs, first principles studies of the $T$ dependence of EFGs are very rare \cite{Das,Toru}.
Several models put forward in the past (see Sec.\ \ref{theories}) rely heavily on phenomenological parameters \cite{Kauf,Haa0}.

In the present paper following the {\it ab initio} path we formulate a novel theoretical approach to this problem,
which on the basis of crucial band structure parameters \cite{lapw}
can describe the $T$ evolution (i.e. decrease or increase) of EFG in detail.
We demonstrate our method by applying it to the hexagonal close packed (hcp) structure of zinc and cadmium
both of which show the characteristic temperature decrease of EFG \cite{Chris,Kauf}.

The paper is written as follows: in the section \ref{sec:model} we give details of our approach,
in section \ref{sec:calc} we describe our results for the $T$ reduction of EFG in hcp zinc and cadmium,
discussing separately the possibility of {\it increasing} EFG with $T$ in Sec.\ \ref{incr}.
Our conclusions are summarized in section \ref{sec:con}.

\section{Theoretical model}
\label{sec:model}

\subsection{Theoretical background and models of the $T$ dependence}
\label{theories}

The EFG tensor $V_{ij}$ is defined as the second partial spatial derivatives of an electric self-consistent-field potential $V$
evaluated at the nuclear site, i.e.
\begin{eqnarray}
    V_{ij} = \frac{\partial^2 V}{\partial i \partial j} ,
\label{i1}
\end{eqnarray}
where $i = x, y, z$. Since $V_{ij}$ is a symmetric (traceless) second rank tensor, it can be further diagonalized
by transforming coordinates to the principal system of axes where $|V_{zz}| \ge |V_{yy}| \ge |V_{xx}|$.
(Thus, the number of independent parameters for EFG in the principal axis system is reduced to two.)
The principal component ($V_{zz}$) is called the electric field gradient, and the second independent parameter
is the asymmetry $\eta$ defined as $\eta = (V_{xx} - V_{yy})/V_{zz}$ ($0 \le \eta \le 1$).

Very often EFGs demonstrate a strong temperature ($T$) dependence which can often be described by a equation of this form:
\begin{eqnarray}
    V_{zz}(T) = V_{zz}(T=0)(1 - B T^{\alpha}) ,
\label{i2}
\end{eqnarray}
where the coefficient $B > 0$ implying smaller EFG with increasing $T$, and the coefficient $\alpha$ is
usually 3/2 \cite{Heu,Chris}.
Later however this ``universal'' form of the $T$ dependence was corrected \cite{Verma}:
it was attributed to normal ($sp$) metals, while for transition metals deviations from the $T^{3/2}$ law were notable
(down to $\alpha \approx 1$). In some cases, the quadratic approximation, $V_{zz}(T) = V_{zz}(0)(1 + C_1 T + C_2 T^2)$, also gave
good quality fits \cite{Verma}.
Even in classical systems like $^{67}$Zn in zinc metal or $^{111}$Cd probes in cadmium metal there were found
deviations from the $T^{3/2}$ law at low temperatures \cite{Verma}.

Since the discovery of strong $T$ dependence of EFG, the problem has been considered theoretically \cite{Jen,Jen2,Nish,Tor,Tho,Toru}.
Two main approaches have been put forward in the past: one is based on electron band structure methods \cite{Jen,Jen2,Tor,Tho},
while the other used screened charged potential formalism \cite{Nish}.
A starting point of both methods is in fact the phenomenological expression for $V_{zz}$ \cite{Wats},
\begin{eqnarray}
    V_{zz} = (1 - \gamma_{\infty}) V_{zz}^{latt} + (1 - R) V_{zz}^{el},
\label{i3}
\end{eqnarray}
where $V_{zz}^{latt}$ is the field gradient due to the non-cubic arrangement of ions in the lattice (excluding the central site),
corrected by the antishielding factor $\gamma_{\infty}$ and $V_{zz}^{el}$ is due to the conduction electrons corrected by the shielding
factor $R$ \cite{Kauf}. The screened charged method works with $V_{zz}^{latt}$ calculating lattice sums over vibrating ions, while
the first approach works with the electron contribution ($V_{zz}^{el}$) and considers $V_{zz}^{latt}$ as virtually
temperature independent. Nowadays however we do not analyze EFG in terms of $V_{zz}^{latt}$, $V_{zz}^{el}$, $\gamma_{\infty}$ and $R$.
With the success of {\it ab initio} all-electron methods for electronic structure of solids \cite{Blaha1,Blaha2,Petr}
capable of treating the electron potential of general shape, the electric field gradient $V_{zz}$ at zero temperature
can be found directly from the obtained self-consistent potential.
Eq.\ (\ref{i3}) then should be rewritten in a general form as
\begin{eqnarray}
    V_{zz} = V_{zz}^{out} + V_{zz}^{in},
\label{i4}
\end{eqnarray}
where $V_{zz}^{in}$ is the local contribution due to EFG, for example, from the charges inside a muffin-tin (MT) sphere,
whereas $V_{zz}^{out}$ is from the charges outside the MT-sphere [more details on Eq.\ (\ref{i4}) are given below in Sec.\ \ref{sub_B}].
Moreover, $V_{zz}^{in}$ and $V_{zz}^{out}$
can be calculated and we know that by far the leading term is $V_{zz}^{in}$ \cite{Toru}.
In our case $V_{zz}^{out}/V_{zz}$ amounts only to -2.7{\%} for Zn and -1.1{\%} for Cd.
(The minus sign of $V_{zz}^{out}$ is discussed in Ref.\ \cite{Haa0}.)

In Ref.\ \cite{Jen} Jena started with $V_{zz}^{in}$ and used reduced matrix elements $M' = M \cdot \exp(-W)$ of pseudopotential
which appeared as a result of averaging $M$ over the lattice vibrations \cite{Kas}. Notice that $\exp(-W)$ is
a square root of the Debye-Waller factor (SRDWF) $\exp(-2W)$. Keeping in $\exp(-W) \approx 1 - W$ only the first order term in $W \ll 1$,
he finally arrived at
\begin{eqnarray}
    V_{zz} = V_{zz}(0)[1 - \beta \varphi(T/T_D)] .
\label{i5}
\end{eqnarray}
Here $T_D$ is an effective Debye temperature and the function $\varphi$ is the Debye integral \cite{Bru}, and $\beta$
is an adjustable constant.
Since at low $T$, $\varphi(T/T_D)$ approaches the zero-point value as $T^2$ and at high $T$ increases linearly with $T$,
in the region of $0 < T/T_D < 2$ a $T^{3/2}$ behavior is approximately followed.
The concept of Ref.\ \cite{Jen} continued in a number of publications on the $T$ dependence of EFG \cite{Jen2,Tor,Tho}.
These studies can not be attributed to a true {\it ab initio} approach, although elements of it are present in Refs.\ \cite{Tho,Kas}.
While the $T$ curves for $V_{zz}$ can be reproduced in some cases, one should be aware that in Eq.\ (\ref{i5})
$T_D$ and especially $\beta$ are fitting parameters of the model. Explicit EFG computation in these models has been avoided in favor
of finding expected trends with temperature.

Nowadays, while we can successfully obtain the $T=0$ value of $V_{zz}$ by performing {\it ab initio} band structure calculations,
the problem of finding its $T$ dependence from first principles remains.
It is a difficult and laborious problem because it requires an accurate calculation of both electronic and phonon properties.
Probably, the first attempt is done by Torumba et al. in Ref.\ \cite{Toru}, who used molecular dynamics and a supercell calculations
of pristine cadmium. For the $T$ dependence of $V_{zz}$ atoms in the supercell were displaced according to the values of mean-square
displacements at this temperature, after which the value of $V_{zz}(T)$ was obtained by averaging over various displacive configurations.
The authors have observed the decrease of EFG with $T$ in the supercell approach, although their data deviate from the experimental
curve at $T>500$~K.

Surprisingly, in some studies an increase of $V_{zz}$ with $T$ has been reported. For example, in 5\%
Fe-doped In$_2$O$_3$ \cite{Sena}, in the rutile modification of TiO$_2$ \cite{Butz13,Butz11}
and in tetrahedrally coordinated Fe sites in Bi$_2$(Fe$_x$Ga$_{1-x}$)$_4$O$_9$ \cite{Web}.
In the rutile structure the situation is in fact probe-dependent:
while EFG measured at $^{47,49}$Ti or substitutional $^{181}$Ta and $^{44}$Sc increases with $T$, it changes in the opposite
direction for $^{111}$Cd probes \cite{Butz11,Butz13}.
The increase of EFG is in disagreement with the classical treatment \cite{Bay,Kus}
of the evolution of EFG in
ionic and molecular crystals based on small rotations of the gradient tensor in respect to a fixed system of axes, according to which
\begin{eqnarray}
    \langle V_{zz} \rangle = V_{zz}(0)\left[1 - \frac{3}{2} \left( \langle \theta_x^2 \rangle + \langle \theta_y^2 \rangle \right) \right] .
\label{i6}
\end{eqnarray}
Here $\theta_x$ and $\theta_y$ are small rotations about the $x-$ and $y-$ axis, respectively, and
$\langle \theta_x \rangle = \langle \theta_y \rangle = 0$.
Notice that Eq.\ (\ref{i6}) always leads to a decrease of EFG with $T$.

Below we formulate a novel theoretical approach to this problem,
which depending on specific conditions can result both
in temperature decrease or increase of EFG with temperature.
It establishes a close relation between the values of the mean-square displacements and the evolution of EFG,
and also uses the square root of the Debye-Waller factor $\exp(-W)$ which lies at the center of early
phenomenological models \cite{Jen,Nish,Kauf}.
The method is applied to the temperature decrease of $V_{zz}$ in hcp zinc and cadmium \cite{Chris,Kauf}.

\subsection{Reduced quadrupole potential and density on the displaced MT-spheres}
\label{sub_A}

The temperature dependence of EFG is clearly a manifestation of both electron and phonon properties.
To evaluate correctly $V_{zz}$ we have to effectively average it over atomic vibrations, because a typical frequency
of the lattice vibrations (1 THz = 10$^{12}$ Hz) is large compared to a typical quadrupole frequency (100 MHz = 10$^8$ Hz)
experienced by nuclear probes in solids.

The electron density of a solid $\rho(\vec{R})$ is a translationally invariant quantity,
\begin{eqnarray}
    \rho(\vec{R} + \vec{a}_i) = \rho(\vec{R}) ,
\label{e1}
\end{eqnarray}
where $\vec{a}_i$ ($i$=1, 2, 3) are the basis vectors of the Bravais lattice.
Therefore, $\rho(\vec{R})$ can be expanded in Fourier series,
\begin{eqnarray}
    \rho(\vec{R}) = \sum_{\vec{K}} \rho(\vec{K}) e^{i \vec{K} \cdot \vec{R}} ,
\label{e2}
\end{eqnarray}
where the vectors $\vec{K}$ belong to the reciprocal lattice.
Fourier components $\rho(\vec{K})$ are usually found quite naturally from solutions
of the Schr{\"o}dinger (or Dirac) equation for an electron in a periodic
mean field potential of the solid.
According to the Bloch theorem, the solution for the electron band $j$ has the form
\begin{eqnarray}
    \Psi_{\vec{k},\,j}(\vec{R}) = \sum_{\vec{K}} c_{\vec{k},\,j}(\vec{K})\, \phi_{\vec{K} + \vec{k}}(\vec{R}) ,
\label{e3}
\end{eqnarray}
where the vector $\vec{k}$ lies in the first Brillouin zone, $\phi_{\vec{K} + \vec{k}}$ are
basis functions and $c_{\vec{k},\,j}(\vec{K})$ are the coefficients of expansion
in the basis set. Almost in all computational methods the basis functions are modified plane waves, which in the
interstitial region are simply $\phi_{\vec{K} + \vec{k}}(\vec{R}) \sim \exp(i(\vec{K} + \vec{k}) \cdot \vec{R})$.
Substituting Eq.\ (\ref{e3}) in
\begin{eqnarray}
    \rho(\vec{R}) = 2 \sum_{\vec{k},j:\, E(\vec{k},j) \le E_F} |\Psi_{\vec{k},\,j}(\vec{R})|^2 ,
\label{e4}
\end{eqnarray}
where $E_F$ is the Fermi energy,
we obtain Eq.\ (\ref{e2}), with the Fourier coefficients $\rho(\vec{K})$ given by
\begin{eqnarray}
    \rho(\vec{K}) = \sum_{\vec{k},j:\, E(\vec{k},j) \le E_F} \sum_{\vec{K}'} c^*_{\vec{k},\,j}(\vec{K}')\, c_{\vec{k},\,j}(\vec{K}' + \vec{K}) .
\label{e5}
\end{eqnarray}

%
\begin{figure}
\resizebox{0.35\textwidth}{!} {
\includegraphics{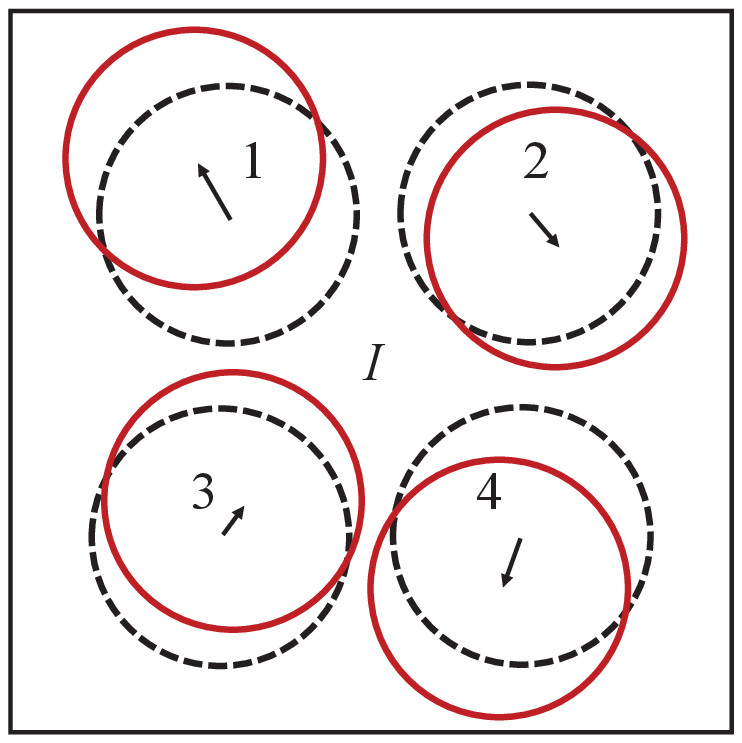}   }

\vspace{2mm}
\caption{
Vibrations of MT-spheres $n = 1-4$ and the interstitial region ($I$).
Dashed lines - averaged surfaces positions in the ideal structure, full (red) lines - instantaneous positions,
arrows indicate the sphere diplacements $u_n$, $n = 1-4$.
As a result of sphere vibrations, the averaged quadrupole potential $\langle V_Q \rangle$ on the sphere surfaces is reduced,
see text for details.
} \label{fig1}
\end{figure}
%
The electron density in the interstitial region is only weakly dependent on the amplitudes of the vibrations
since the band electron on average experiences a periodic potential from the regular arrangement of the ions.
On the other hand, the electron density around a nucleus being attached to it by the Coulomb interaction
is expected to follow adiabatically the vibrating ion.
To reconcile these apparently contradictory viewpoints we consider a model of vibrating spheres
[called below muffin-tin or MT-spheres], Fig.\ \ref{fig1}, which are immersed in the crystal.
We keep the values of $\rho(\vec{K})$, Eq.\ (\ref{e5}) in the interstitial region unchanged while the electron
density on the vibrating sphere will be calculated taking into account the nuclear displacements.
It turns out that at a finite and even zero temperature $T$ the quantities $\rho(\vec{K})$ on vibrating MT-spheres are effectively reduced (see below).
In the language of the LAPW method this implies a modification of boundary conditions for the solutions inside the MT-sphere \cite{flapw},
which in turn changes EFG.
Therefore, our main goal is to calculate the temperature dependence of the electron density on the vibrating sphere,
which later (Sec.\ \ref{sub_C}) will be used for the calculation of EFG.
In the following we consider a simple case of a single atom in the primitive unit cell.
(This avoids unnecessary technical complications whereas
the generalization for the case of few atoms is straightforward.)

The electron density on an MT-sphere of the radius $R_{MT}$ centered at a site $n$
and displaced from the equilibrium position $\vec{X}(n) =\left\{ X_x(n),X_y(n),X_z(n) \right\}$ by the vector $\vec{u}_{n}$
can be expanded in the multipole series
\begin{eqnarray}
    \rho(\vec{u}_n,\, r = R_{MT}, \theta, \phi) = \sum_{\Lambda} \rho_{\Lambda}(\vec{u}_n)\, S_{\Lambda}(\theta, \phi) ,
\label{e6}
\end{eqnarray}
where $\Lambda \equiv (l, \tau)$ stands for $A_{1g}$ irreducible representations of the crystal point group,
$S_{\Lambda}(\theta, \phi)$ are corresponding symmetry adapted functions (SAFs),
$l$ is the multipole orbital index and $\tau$ counts functions with the same $l$ (if there are few such functions).
The polar angles $\Omega = (\theta, \phi)$ are specified by the vector $\vec{r}$ from the nuclear position ($\vec{X} + \vec{u}$).
$S_{\Lambda}(\theta, \phi)$ are linear combinations of real spherical harmonics ($Y_l^{m,c} \sim \cos m\phi$ and $Y_l^{m,s} \sim \sin m\phi$)
defined by the crystal site symmetry and tabulated in \cite{BC}.
The first function with $\Lambda = 0$ ($l=0$, $\tau=1$) is the spherically symmetric (monopole) contribution, $S_0 = Y_{l=0}^{m = 0} = 1/\sqrt{4 \pi}$.
For the hexagonal close packed structure defined by the basis vectors specified in Ref.\ \cite{BC} the other
multipole functions are: $Y_{l=2}^{m=0}$, $Y_{l=3}^{3,c}$, $Y_{l=4}^{m=0}$, $Y_{l=5}^{3,c}$, $Y_{l=6}^{m=0}$, $Y_{l=6}^{6,c}$, etc.
The index $\Lambda$ runs over $(0,1) \equiv 0$, $(2,1) \equiv Q$, (3,1), (4,1), (5,1), (6,1), (6,2), etc.
Thus, we have only one quadrupolar function $S_Q \equiv S_{l=2}=Y_{l=2}^{m=0}$, which explicitly reads as
\begin{eqnarray}
    S_Q (z) = \frac{1}{4} \sqrt{\frac{5}{\pi}} (3z^2 - 1) ,
\label{e6b}
\end{eqnarray}
where $z = \cos \theta$.
Then the general equation (\ref{e6}) can be written as
\begin{eqnarray}
    \rho(\vec{u},\, \theta, \phi) = \frac{ \rho_{0}(\vec{u})}{\sqrt{4 \pi}} + \rho_{Q}(\vec{u})\, S_Q(\theta, \phi) \nonumber \\
    + \rho_{(3,1)}(\vec{u})\, Y_{l=3}^{3,c}(\theta, \phi) + ... \;.
\label{e6c}
\end{eqnarray}

Rewriting the plane wave expansion in spherical harmonics [e.g. Eq.\ (34.3) of Ref.\ \cite{Lan}] in the complete basis set of
symmetry adapted spherical harmonics [Eq.\ (2.4) of Ref.\ \cite{Nik}] one can obtain the expansion in terms of SAFs
centered at the displaced nucleus of the site $n$ given by $\vec{X}(n) + \vec{u}_n$,
\begin{eqnarray}
     e^{i \vec{K} \cdot \vec{R}} &=& e^{i \vec{K} (\vec{X}(n) + \vec{u}_n )}\, 4 \pi \nonumber \\
      & & \sum_{\Lambda} i^l j_l(K r(n)) S_{\Lambda}(\hat{K})\, S_{\Lambda}(\theta(n), \phi(n)) .
\label{e7a}
\end{eqnarray}
Here $j_l$ are spherical Bessel functions and $\hat{K}$ specifies the direction of $\vec{K}$, i.e. $\hat{K} \equiv (\theta_{\vec{K}},\phi_{\vec{K}})$.
From Eq.\ (\ref{e2}) then we obtain for the coefficients $\rho_{\Lambda}$, Eqs.\ (\ref{e6}), (\ref{e6c}), at any chosen site $n$:
\begin{eqnarray}
     \rho_{\Lambda}(\vec{u}) = 4 \pi i^l \sum_{\vec{K}} j_l(K R_{MT}) S_{\Lambda}(\hat{K})\, \rho_{\vec{u}}(\vec{K}) ,
\label{e7}
\end{eqnarray}
where $\vec{u} \equiv \vec{u}_n$ and
\begin{eqnarray}
     \rho_{\vec{u}}(\vec{K}) = e^{i \vec{K} \cdot \vec{u}} \rho(\vec{K}) .
\label{e9}
\end{eqnarray}
Here we have taken into account that for vectors $\vec{K}$ and $\vec{X}(n)$ belonging to the reciprocal and direct lattice respectively,
$e^{i \vec{K} \vec{X}} = 1$.
In the case $\vec{u} = 0$ (i.e. at the equilibrium position $\vec{X}$) we get
$\rho_{\vec{u}}(\vec{K}) = \rho(\vec{K})$,
where $\rho(\vec{K})$ is defined by Eq.\ (\ref{e5}).
Averaging over the displacements at the chosen site $n$ we arrive at
\begin{eqnarray}
    \langle \rho_{\Lambda} \rangle = \langle \rho_{\Lambda}(\vec{u}) \rangle = 4 \pi i^l \sum_{\vec{K}} j_l(K R_{MT}) S_{\Lambda}(\hat{K})\,
     \langle \rho(\vec{K}) \rangle , \nonumber \\
\label{e10}
\end{eqnarray}
where
\begin{eqnarray}
    \langle \rho(\vec{K}) \rangle = \langle e^{i \vec{K} \cdot \vec{u}} \rangle \rho(\vec{K})  .
\label{e11}
\end{eqnarray}
Notice that Eqs.\ (\ref{e10}) and (\ref{e11}) are independent of $n$ because the thermal averages
$\langle \rho(\vec{K}) \rangle \equiv \langle \rho_{\vec{u}_n}(\vec{K}) \rangle$
and $\langle \exp (i \vec{K} \cdot \vec{u}_n) \rangle$ are the same for all equivalent atoms.
As has been proved by Glauber \cite{Gla} the thermal average on the right hand side of Eq.\ (\ref{e11})
can be transformed to
\begin{eqnarray}
    \langle e^{i \vec{K} \cdot \vec{u}} \rangle = e^{-W(\vec{K},T)}  ,
\label{e12}
\end{eqnarray}
where 
\begin{eqnarray}
    W(\vec{K},T) = \frac{1}{2} \langle (\vec{K} \cdot u)^2 \rangle .
\label{e12'}
\end{eqnarray}
Since the usual Debye-Waller factor is $\exp(-2W)$ \cite{Bru}, the temperature function $\exp(-W)$, Eq.\ (\ref{e12}),
is the square root of the Debye-Waller factor (SRDWF).
Such a function has been used by Kasowski
for the description of the temperature-dependent Knight shift in cadmium \cite{Kas}.
The function $W$ can also be written in $k-$space \cite{Bru} in the familiar form
\begin{eqnarray}
    W(\vec{K},T) = \frac{1}{2 N} \sum_{\vec{k},s} \langle (\vec{K} \cdot u_s(\vec{k}))^2 \rangle ,
\label{e14}
\end{eqnarray}
where the summation runs over all vectors $\vec{k}$ in the first Brillouin zone and the phonon branches $s$,
while $u_s(\vec{k})$ stands for the corresponding phonon amplitudes.
From space symmetry considerations it follows that the function $W(\vec{K})$ is the same
for a set of (non-equivalent) vectors (rays) $\vec{K}_i$ obtained from $\vec{K}_1 = \vec{K}$ by
the application to $\vec{K}_1$ all rotational or mirror symmetry elements
of the crystal point group (i.e. for the vectors $\vec{K}_i$ belonging to the same star).
In the harmonic approximation $W$ implicitly depends on the temperature through the number of thermally excited phonons $n_{\vec{K},s}$.
In the Debye approximation at high temperatures ($T \gg T_D$) $W \sim T$.

For the calculation of EFG we need the quadrupole component ($\Lambda = (2,1)$) of the electron density and potential, Appendix~A.
For that purpose, with the final expressions (\ref{e10}), (\ref{e11}), (\ref{e12}) and (\ref{e12'}) we obtain
for the average quadrupole component $\langle \rho_{Q} \rangle$ of density on the MT-sphere in Eq.\ (\ref{e6c}),
\begin{eqnarray}
    \langle \rho_{Q} \rangle = -4 \pi \sum_{\vec{K}} j_2(K R_{MT}) S_{Q}(\hat{K})\, e^{-W(\vec{K},T)}\, \rho(\vec{K}) .
\quad \label{e15}
\end{eqnarray}
This expression will be used later in Sec.\ \ref{sub_C}.

Since for all $\vec{K} \neq 0$ we have $W(\vec{K}) > 0$ even for zero temperature,
the corresponding averaged Fourier components $\langle \rho(\vec{K}) \rangle$
in Eqs.\ (\ref{e10}) and (\ref{e11}) are effectively reduced.
The reduction means that as a rule the quadrupole component ($\langle \rho_Q \rangle$ in Eq.\ (\ref{e6c}))
and also the other components with $\Lambda' > 0$ (i.e. $\langle \rho_{(3,1)} \rangle$ etc.)
on the MT-sphere surface are decreased in absolute value in comparison with their static values:
$|\langle \rho_{\Lambda'}(\vec{u}) \rangle | < |\rho_{\Lambda'}(\vec{u}=0)|$.
As will be discussed below in Sec.\ \ref{sub_C} the effective decrease of $\langle \rho_Q \rangle$
occurs also in the interior of the MT-sphere, $r \leq R_{MT}$, and finally leads to a reduction of EFG.
In some rare cases as a result of the interplay between terms of opposite signs in Eq.\ (\ref{e15})
an increase of $\langle \rho_Q \rangle$ can occur. Such a situation is discussed in detail below in Sec.~\ref{incr}.

The reduction or increase of $\langle \rho_Q \rangle$ however does not affect the total charge of the sphere
because the integral over the polar angles for SAFs $S_Q$, $Y_3^{3,c}$ and the others with $l \neq 0$ are zero.
The total charge of the sphere is due to the monopole term ($l = 0$),
which is closely related to the $\vec{K} = 0$ term in the Fourier expansion of density, Eq.\ (\ref{e2}).
Notice that the $\rho(K = 0)$ component is independent of $T$, Eq.\ (\ref{e11}),
keeping the total charge inside the sphere approximately constant.
A very small decrease of the average value of $\langle \rho_0 \rangle$ in Eq.\ (\ref{e6c})
(implying a small increase of the charge in the interstitial region) can be
accounted for by a small increase of the $\rho(K = 0)$ component of
the Fourier expansion (\ref{e2}).
In our calculations the effect is found small and has been neglected.

In the following we shall apply our approach to the case of the hexagonal close packed structure,
although in principle the consideration is general and can be used for other non-cubic lattices.

\subsection{EFG and quadrupole ($L=2$) expansion of the electron potential around the nucleus}
\label{sub_B}

In this section we demonstrate that the reduction (or increase) of $\langle \rho_Q \rangle$ inside the MT-sphere
results in a corresponding change of the quadrupole component of the potential and the electric field gradient.

\begin{figure}[!]
\resizebox{0.35\textwidth}{!}
{\includegraphics{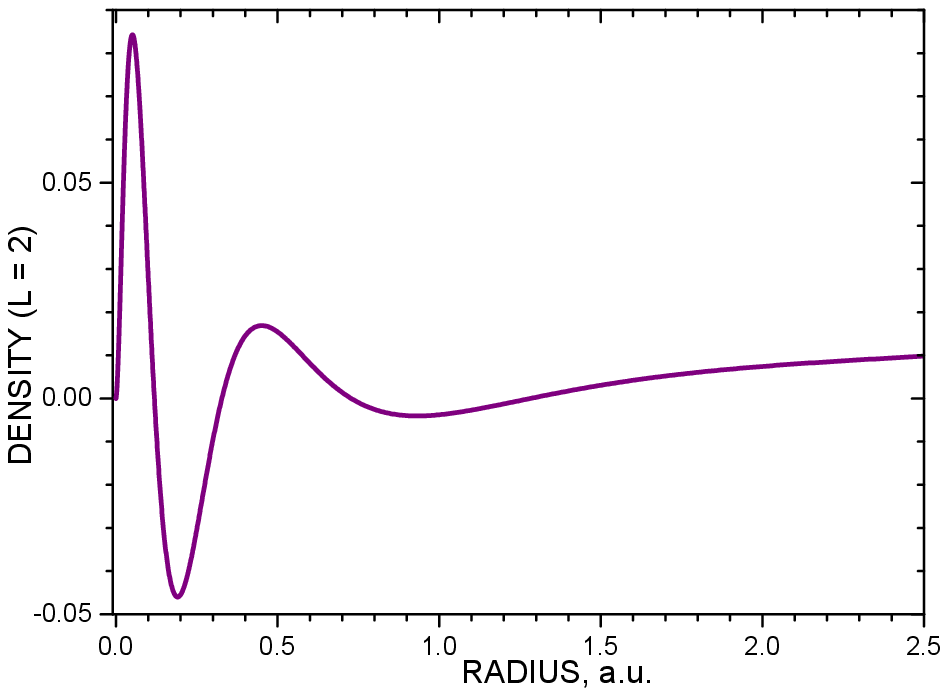}}
\vspace{2mm}
\resizebox{0.35\textwidth}{!}
{\includegraphics{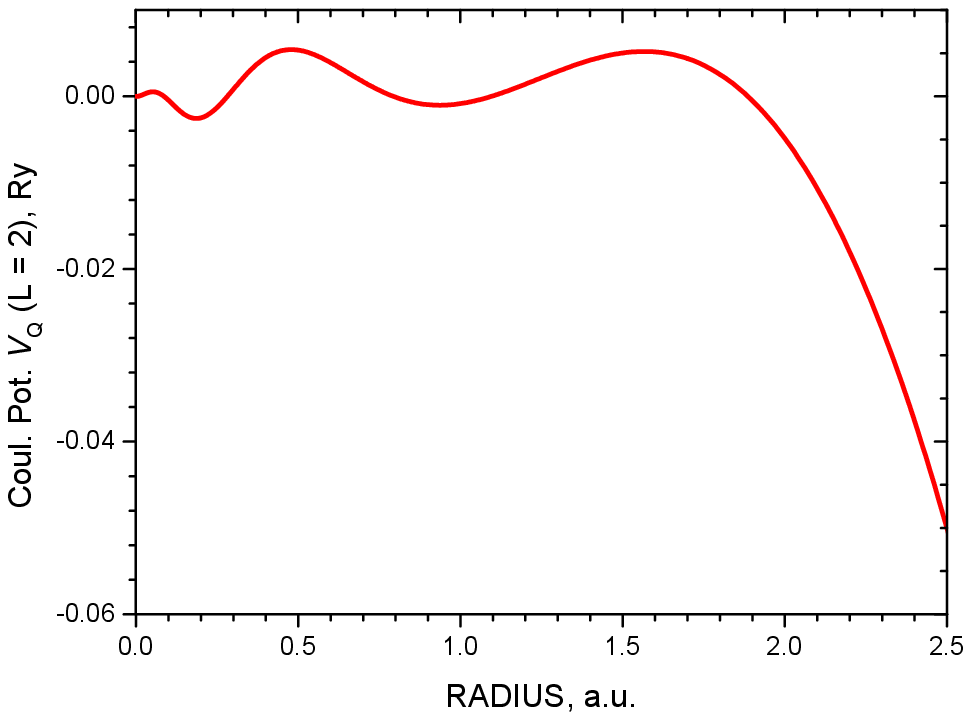}}

\vspace{2mm}
\caption{
DFT calculation of the quadrupole ($l = 2$) component $\rho_Q(r)$ of electron density (top panel)
and the corresponding Coulomb potential $V_Q(r)$ (bottom panel) inside the MT-sphere of the hcp structure of Zn
($a=2.659$ {\AA}, $c=4.851$ {\AA}, $R_{MT}=1.33$ {\AA} or 2.51 a.u.).
} \label{fig2}
\end{figure}

\begin{figure}[!]
\resizebox{0.35\textwidth}{!}
{\includegraphics{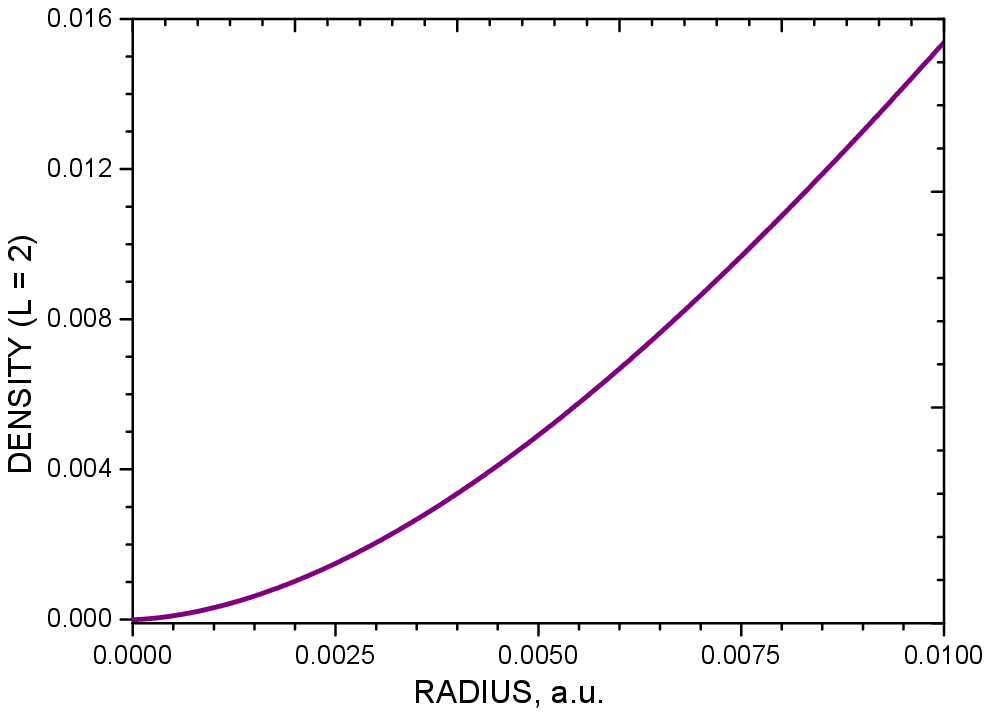}}
\vspace{2mm}
\resizebox{0.36\textwidth}{!}
{\includegraphics{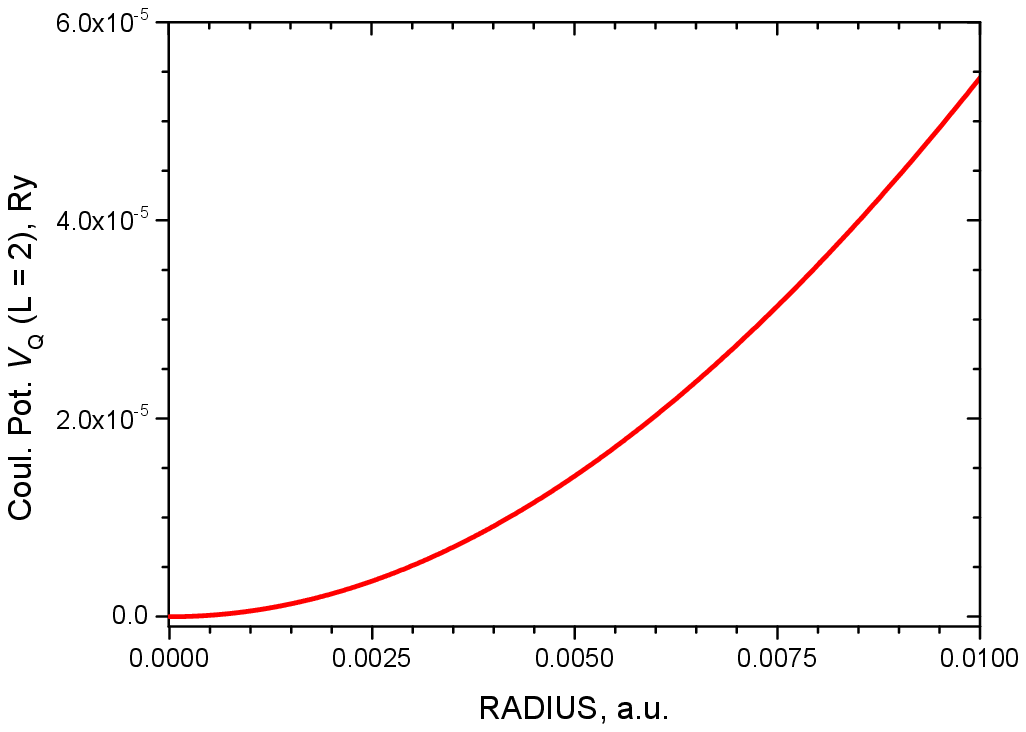}}

\vspace{2mm}
\caption{
DFT calculation of the quadrupole ($l = 2$) component $\rho_Q(r)$ of electron density (top panel)
and the corresponding Coulomb potential $V_Q(r)$ (bottom panel) of the hcp structure of Zn in the vicinity
of nucleus. (Fig.\ \ref{fig3} zoomed at $r=0$.)
Both the density and potential are proportional to $r^2$.
For comparison, the Zn nuclear radius is $9.1 \cdot 10^{-5}$ a.u.
} \label{fig3}
\end{figure}

Similarly to the density expansion, Eq.\ (\ref{e6}),
the electron self-consistent-field Coulomb potential of the general form around a nucleus, employed for example in the full potential
linearized augmented plane wave method (FLAPW) \cite{flapw,wien2k}, also can be expanded in multipolar series,
\begin{eqnarray}
    V(r, \theta, \phi) = \sum_{\Lambda} V_{\Lambda}(r)\, S_{\Lambda}(\theta,\phi) .
\label{p1}
\end{eqnarray}
Notice that the coefficients $V_{\Lambda}(r)$ for the potential and $\rho_{\Lambda}(r)$ for electron density, Eq. (\ref{e6}),
can be obtained from {\it ab initio} electron band structure calculations.
In Figs.\ \ref{fig2} and \ref{fig3} we reproduce such dependencies for the quadrupole component, $l=2$.

From a scrupulous analysis of the multipolar components of the potential (see Appendix~A and Ref.\ \cite{Nik})
it follows that in the neighborhood of the nucleus (when $r \ll 1$) the radial dependence of the function $V_{l,\tau}$ reads as
\begin{eqnarray}
  V_{\Lambda}(r) = v_{\Lambda}\, r^l ,
\label{p2}
\end{eqnarray}
where $v_{\Lambda}$ is a constant. This in particular
implies that for quadrupolar component [$\Lambda = (2,1)$] we have $V_{Q} \equiv V_{(2,1)}(r) = v_{Q}\, r^2$, for the component $\Lambda = (4,1)$
 $V_{(4,1)}(r) = v_{(4,1)}\, r^4$, etc. This dependence of $V_{Q}(r)$ is illustrated in Fig.\ \ref{fig3}.
It is worth noting that the quadrupolar electron density component also follows the same law close to the nucleus, i.e.
$\rho_Q(r) \propto r^2$, Fig.\ \ref{fig3}.

%
%

%
%

Taking Eq. (\ref{p2}) into account, we obtain for the electric field gradient,
\begin{eqnarray}
  V_{zz} = \frac{ \partial^2 V_{Q}}{\partial z^2} = \sqrt{\frac{5}{\pi}} \, v_{Q} .
  \label{p5}
\end{eqnarray}
Details are given in Appendix \ref{append_A}.
Notice that Eq.\ (\ref{p5}) gives a temperature independent EFG.

The quadrupole potential on the MT-sphere, $V_Q^S$ consists of two
contributions,
\begin{eqnarray}
  V_Q^S = V_Q^{S,\, out} + V_Q^{S,\, in} .
  \label{p6}
\end{eqnarray}
Here, $V_Q^{S,\, out}$ and $V_Q^{S,\, in}$ are the potentials due to all charges outside
the MT-sphere and inside it, respectively.
Correspondingly, for the quadrupole potential $V_Q$ at any point $\vec{r} = (r,\, \theta,\phi)$ inside the MT-sphere we have
\begin{subequations}
\begin{eqnarray}
  V_Q(\vec{r}) = V_Q^{out}(\vec{r}) + V_Q^{in}(\vec{r}) ,
  \label{p7a}
\end{eqnarray}
where
\begin{eqnarray}
  V_Q^{out}(r, \theta,\phi) = V_Q^{S,\, out} \frac{r^2}{R_{MT}^2}\, S_Q(\theta,\phi) ,
  \label{p7b}
\end{eqnarray}
(for hcp lattices $S_Q = Y_2^0$) and
\begin{eqnarray}
  V_Q^{in}(r, \theta,\phi) = \frac{4 \pi}{5} \left( \frac{q_Q(r)}{r^3} + r^2 q'_Q(r)  \right) S_Q(\theta,\phi) .
  \label{p7c}
\end{eqnarray}
\end{subequations}
Here
\begin{subequations}
\begin{eqnarray}
   & & q_Q(r) = \int_0^r \rho_Q(r')\, {r'}^3 dr',
  \label{p8a} \\
   & & q'_Q(r) = \int_r^{R_{MT}} \frac{\rho_Q(r')}{r'} dr',
  \label{p8b}
\end{eqnarray}
\end{subequations}
Since $q'_Q(R_{MT})=0$, we find $V_Q^{S,\, in}=4\pi q(R_{MT})/R_{MT}^3$.
Notice that $V_Q^{out}$ is explicitly proportional to $r^2$, Eq.\ (\ref{p7b}).
It can be shown \cite{Nik} that the same holds for $V_Q^{in}$ at $r \ll 1$.
From Eqs.\ (\ref{p2}), (\ref{p5}) we then arrive at
\begin{eqnarray}
  V_{zz} = \sqrt{\frac{5}{\pi}} \left( v_{Q}^{out} +  v_{Q}^{in} \right).
  \label{p9}
\end{eqnarray}
Here $v_{Q}^{out}$ and $v_{Q}^{in}$ are obtained from $V_Q^{out}/r^2$ and $V_Q^{in}/r^2$
when $r \rightarrow 0$.
As a result of the reduction of $\langle \rho_Q(r) \rangle$ with temperature, $\langle v_{Q}^{out} \rangle$ and $\langle v_{Q}^{in} \rangle$ also become reduced,
and $\langle V_{zz} \rangle$ decreases with temperature.

\subsection{Temperature dependence of EFG}
\label{sub_C}

As shown in Sec.\ \ref{sub_A} the average quadrupole density component $\langle \rho_Q(R_{MT}) \rangle$
on the vibrating MT-sphere as a rule is reduced, Eq.\ (\ref{e15}).
However, to calculate exactly the effect of the reduction on EFG, we have to know $\langle \rho_Q(r) \rangle$ at all values $r \le R_{MT}$,
see Sec.~\ref{sub_B}.
In a straightforward approach we should perform a calculation for $\langle \rho_Q(r) \rangle$ inside the MT-sphere considering the new value of
$\langle \rho_Q(R_{MT}) \rangle$, Eq.\ (\ref{e15}),
as an input boundary condition.
In practice, by changing $R_{MT}$ in a range close to the maximal (contact) radius $R_{max}$,
we have found that the reduction of the quadrupolar density component,
$\rho_Q(r,T)/\rho_Q(r,\,T=0)$, is approximately constant.
In Zn for example the change of $R_{MT}$ from the $R_{max}$ value of
2.517 a.u. to 2.30 a.u. at 300 K results in a change of the ratio from 0.945 to 0.926 (2{\%}).
We then in a first approximation assume that for all $r \le R_{MT}$
\begin{eqnarray}
   \frac{\rho_Q(r,T)}{\rho_Q(r,\,T=0)} = \frac{\rho_Q(R_{MT},T)}{\rho_Q(R_{MT},\,T=0)} = R^{in}(T) .
  \label{t1}
\end{eqnarray}

Accoring to Eqs.\ (\ref{p8a}) and (\ref{p8b}), the reduced quadrupolar density component $\langle \rho_Q(r,T) \rangle$ for all $r \le R_{MT}$ changes
the quadrupole charges as $q_Q(r,T)=R^{in}(T)\, q_Q(r)$ and $q'_Q(r,T)=R^{in}(T)\, q'_Q(r)$.
This in turn leads to the reduction of $V_Q^{in}(T)=R^{in}(T)\, V_Q^{in}$, Eq.\ (\ref{p7c}), and $v_{(2,0)}^{in}(T) = R^{in}(T)\, v_{(2,0)}^{in}$.
We then for the temperature evolution of EFG have
\begin{eqnarray}
  V_{zz}(T) = \sqrt{\frac{5}{\pi}} \left( R^{out}(T)\, v_{(2,0)}^{out} +  R^{in}(T)\, v_{(2,0)}^{in} \right) \nonumber \\
  \label{t2}
\end{eqnarray}
[compare with Eq.\ (\ref{p9})].
Here the factor $R^{out}(T)$ accounts for the change of the potential due to all charges outside the MT-sphere,
while $v_{(2,0)}^{out}$ and $v_{(2,0)}^{in}$ are temperature independent (calculated with an {\it ab initio} electron band structure method).
In practice, $v_{(2,0)}^{out}$ is found to be small (1-3 {\%}) compared to $v_{(2,0)}^{in}$.
Since, in addition the difference between $R^{out}$ and $R^{in}$ is not essential, in the following for simplicity we take $R^{out} \approx R^{in}$.
Then the temperature dependence of EFG is completely due to the change of $\rho_Q$ inside
the MT-sphere and
\begin{eqnarray}
   V_{zz}(T) \approx R^{in}(T) V_{zz} .
  \label{t3}
\end{eqnarray}
Eqs.\ (\ref{t3}) and (\ref{t1}) imply that we consider the change of the quadrupole interaction as a perturbation
causing a linear effect inside the MT-sphere.

\subsection{Mean-square displacements}
\label{sub_D}

\begin{figure}[!]
\resizebox{0.45\textwidth}{!}
{\includegraphics{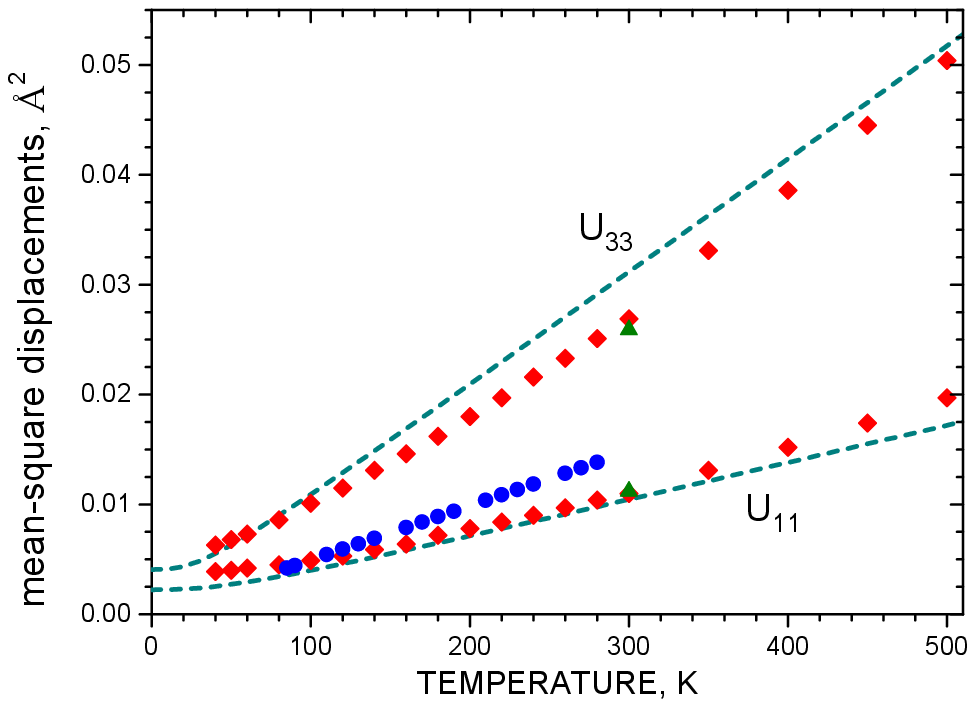}
}

\vspace{2mm}
\caption{
Temperature evolution of $U_{11}= \langle u_{x}^2 \rangle$ and $U_{33}= \langle u_{z}^2 \rangle$ for hcp Zn
found by various methods.
Dark cyan dash lines stand for calculations in the harmonic approximation (DFPT of QE \cite{QE1,QE2}),
red diamonds are experimental data from the single crystal structural refinement of Zn at different $T$ \cite{Nus},
olive triangles are experimental data at $T=300$~K \cite{Ske},
blue circles are obtained from experimentally determined phonon density of states \cite{Pen}.
} \label{fig4}
\end{figure}

The SRDW factor, Eqs.\ (\ref{e12}) and (\ref{e12'}) depends crucially on the mean square
displacements $\langle u_x^2 \rangle$, $\langle u_y^2 \rangle$ and $\langle u_z^2 \rangle$,
which are functions of temperature.
For the hcp lattice $\langle u_x^2 \rangle = \langle u_y^2 \rangle$, and for calculations
of SRDWF we need to know only two functions: $U_{11}(T) = \langle u_x^2 \rangle$
and $U_{33}(T) = \langle u_z^2 \rangle$ (for hcp lattice $U_{33} > U_{11} $),
\begin{eqnarray}
   W(\vec{K}, T) = \frac{1}{2} \left( (K_x^2 + K_y^2)\, U_{11}(T) + K_z^2\, U_{33}(T)  \right) .
  \label{m1}
\end{eqnarray}
The functions $U_{11}(T)$ and $U_{33}(T)$ can be calculated within the harmonic approximation
or extracted experimentally as illustrated in Fig.\ \ref{fig4} for the hcp lattice of Zn.
%
%

\begin{figure}[!]
\resizebox{0.45\textwidth}{!}
{\includegraphics{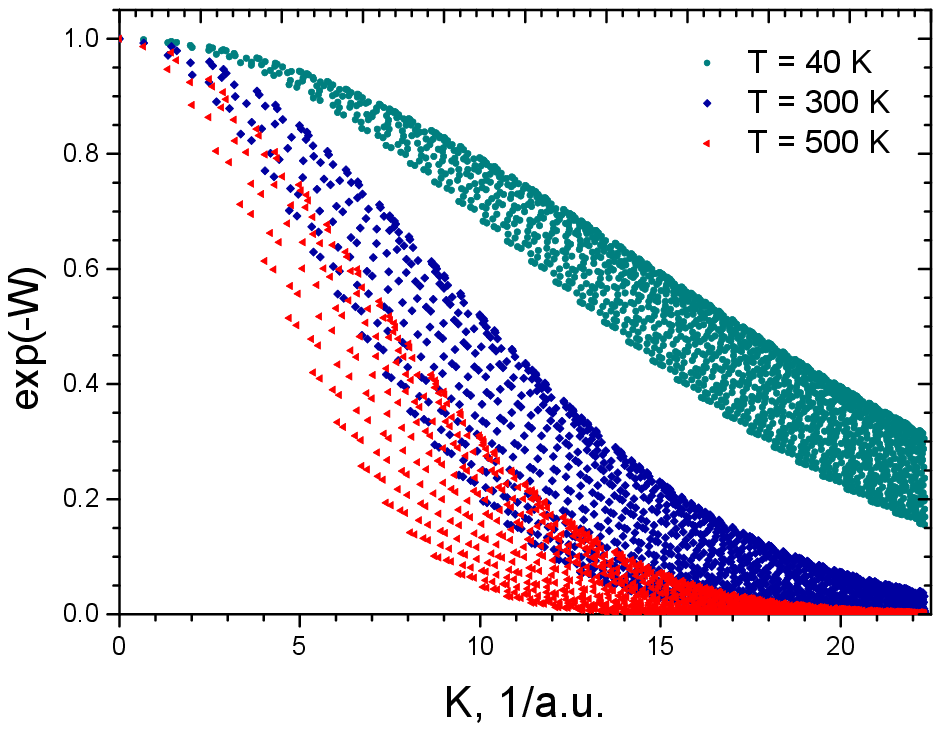}
}

\vspace{2mm}
\caption{
Square root of the Debye-Waller factors (SRDWF), $\exp(-W(\vec{K},\, T))$, as functions of
the modulus of reciprocal vectors $K = |\vec{K}|$ of hcp Zn for various temperatures.
Dark cyan circles are data for $T = 40$ K, blue squares - $T = 300$ K and red triangles - $T = 500$ K.
Calculations are based on experimental data for the mean-square displacements $U_{11}(T)$ and $U_{33}(T)$, Ref.\ \cite{Nus}.
} \label{fig5}
\end{figure}

In Fig.\ \ref{fig5} we plot individual SRDW factors, $\exp(-W(\vec{K}, T))$, for the hcp structure of Zn,
which are responsible for the temperature reduction of the quadrupole density and potential, Sec.\ \ref{sub_A}.
With the increase of $T$ the SRDW factors get shifted to lower values.
A finite width of SRDWF for reciprocal vectors with close values of $|\vec{K}|$ is due to different orientations of $\vec{K}$ in respect to
the ellipsoid of mean-square displacements. The width becomes larger with the increase of asymmetry between $U_{11}$ and $U_{33}$.
%
%

\section{Application to zinc and cadmium}
\label{sec:calc}

In this section we apply the method described earlier
to calculations of the temperature dependence of the electric field gradient
of solid zinc and cadmium crystallized in the hexagonal close packed
lattice.

For that we need
data on the Fourier expansion of the electron density in the interstitial region,
quadrupole components of electron density and potential obtained from the electron band structure calculations.
Electron density functional calculations have been performed with the Moscow-FLAPW code \cite{lapw}.
The code explicitly takes into account the nuclear size and the change of the potential and the wave function inside
the nuclear region to obtain the electric field gradient accurately. In addition, the number of radial points
inside the MT-region has been increased to 3000 (for some runs 3500).
The typical LAPW basis set cut-off parameter was $R_{MT} K_{max}$ = 9 (the number of basis functions 221),
the number of $k$ points was 704 (1331), the maximal value of the LAPW plane wave expansion $l_{max}=10$.
We have used the tetrahedron method for the linear interpolation of energy between $k$ points \cite{tet}.
For calculation of the exchange-correlation
potential and the exchange-correlation energy contribution we
have used the Perdew-Burke-Ernzerhof (PBE) variant \cite{PBE} of the generalized-gradient approximation (GGA) within
the density functional theory (DFT) at
experimental lattice constants at room temperature
[$a=2.663$ {\AA}, $c=4.944$ {\AA}, $R_{MT}=1.196$ {\AA} (Zn) and
$a=2.996$ {\AA}, $c=5.674$ {\AA}, $R_{MT}=1.429$ {\AA} (Cd).]

Also, the temperature evolution of the tensor of mean-square displacements,
$U_{11}(T)$ and $U_{33}(T)$, is
required for the calculation of the square root of the Debye-Waller factor, $\exp(-W(\vec{K}, T))$, for vectors $\vec{K}$
of the reciprocal lattice, Eqs.\ (\ref{m1}) and (\ref{e12}).
The temperature evolution of the tensor of mean-square displacements
can be obtained by three different ways:
(1) from a direct experimental parametrization of the displacements $\langle u_{x}^2(T) \rangle$ and $\langle u_{z}^2(T) \rangle$ as
given in Ref.\ \cite{Nus} for Zn;
(2) from calculations of the phonon frequencies and eigenvectors
(for example, within the DF perturbation treatment of lattice dynamics with Quantum Espresso \cite{QE1,QE2});
(3) from effective Debye temperature $T_D(T)$ which appear as a result of experimental parametrization
\cite{Kri,Pen}. The strict harmonic approximation is adopted only in (2), while in (1) and (3)
an anharmonic effects such as thermal expansion of solids and phonon softening are effectively taken into account.
Experimentally, the values of $\langle u_{x}^2(T) \rangle$ and $\langle u_{z}^2(T) \rangle$
(or directly related to them temperature factors $B_x(T)$ and $B_z(T)$ \cite{Pen}) are found from the
temperature evolution of the x-ray diffraction spectra.

For the phonon part (2) of calculations
the first-principles pseudopotential method as implemented in the Quantum Espresso (QE) package \cite{QE1,QE2}
with the PBE exchange-correlation functional \cite{PBE} has been employed.
The projected-augmented-wave (PAW) type scalar-relativistic pseudopotentials have been taken from the standard solid-state pseudopotential (SSSP)
library \cite{sssp1,sssp2}.
The integration over the Brillouin zone (BZ) for the electron density of states has been performed on a $24 \times 24 \times 12$ grid of $k$-points,
the plane-wave kinetic cut-off energy was 70 Ry.
The lattice-dynamical calculations have been carried out within the density-functional perturbation theory (DFPT).
Phonon dispersions have been computed using the interatomic force constants based on a $6 \times 6 \times 4$ $k$-point grid,
with a $48 \times 48 \times 24$ grid used to obtain the phonon densities of states and $U_{11}(T)$ and $U_{33}(T)$.

\subsection{Zinc}
\label{Zn}

\begin{figure}[!]
\resizebox{0.45\textwidth}{!}
{\includegraphics{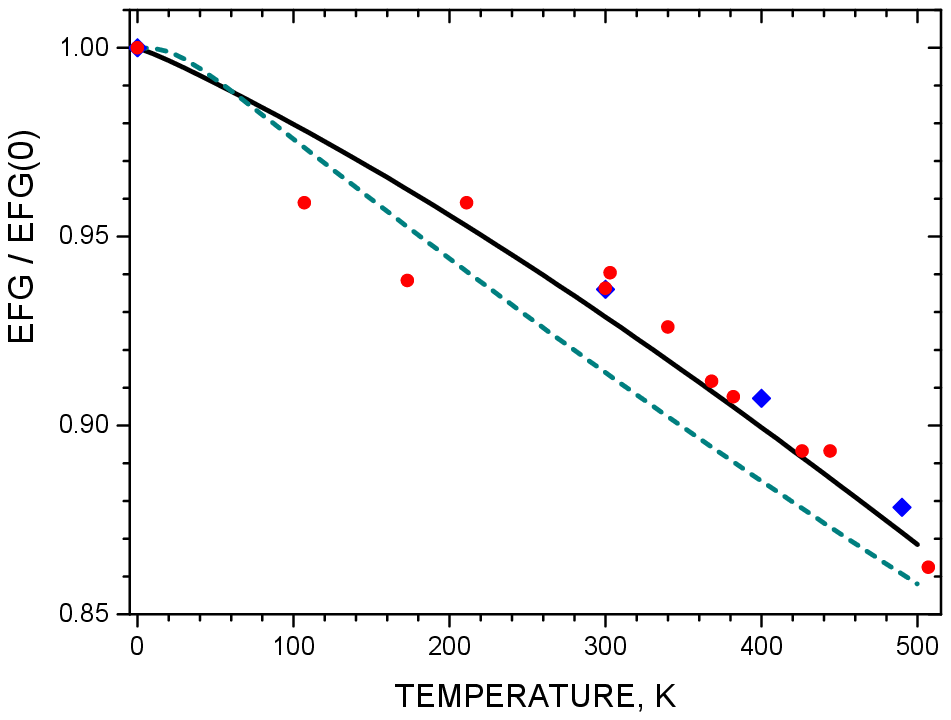}
}

\vspace{2mm}
\caption{
Temperature dependence of electric field gradient in Zn, $V_{zz}(T)/V_{zz}(T=0)$.
Calculations -- solid black line -- are based on experimental data for the mean-square displacements $U_{11}(T)$ and $U_{33}(T)$, Ref.\ \cite{Nus}.
DFPT-QE calculations \cite{QE1,QE2} -- dark cyan dash line -- have been made in the harmonic approximation.
Red circles \cite{Chris} and blue diamonds \cite{Bertschat} stand for experimental data on EFG ratios.
} \label{fig6}
\end{figure}

Mean-square displacements for hcp structure of zinc are shown in Fig.\ \ref{fig4}.
First, we notice that in the DFPT-QE harmonic approximation
in the region of $T$ from 60 K to 500~K, $U_{11}(T)$ and $U_{33}(T)$
 are practically linear in $T$.
Such behaviour is also expected in the Debye model for $T > T_D$.
Experimentally, however we observe that $U_{11}(T)$ and $U_{33}(T)$ deviate from the
linear law. As shown in Ref.\ \cite{Nus} $U_{11}(T)$ and $U_{33}(T)$ are approximated
by quadratic functions of $T$ for all data in the range from 40 K to 500 K.
In addition, in the harmonic approximation values of $U_{33}$ are
slightly overestimated, Fig.\ \ref{fig4}, which results in a considerable suppression
of the SRDF factors and the calculated temperature dependence of EFGs, Fig.\ \ref{fig6}.

%
\begin{figure}[!]
\resizebox{0.35\textwidth}{!}
{\includegraphics{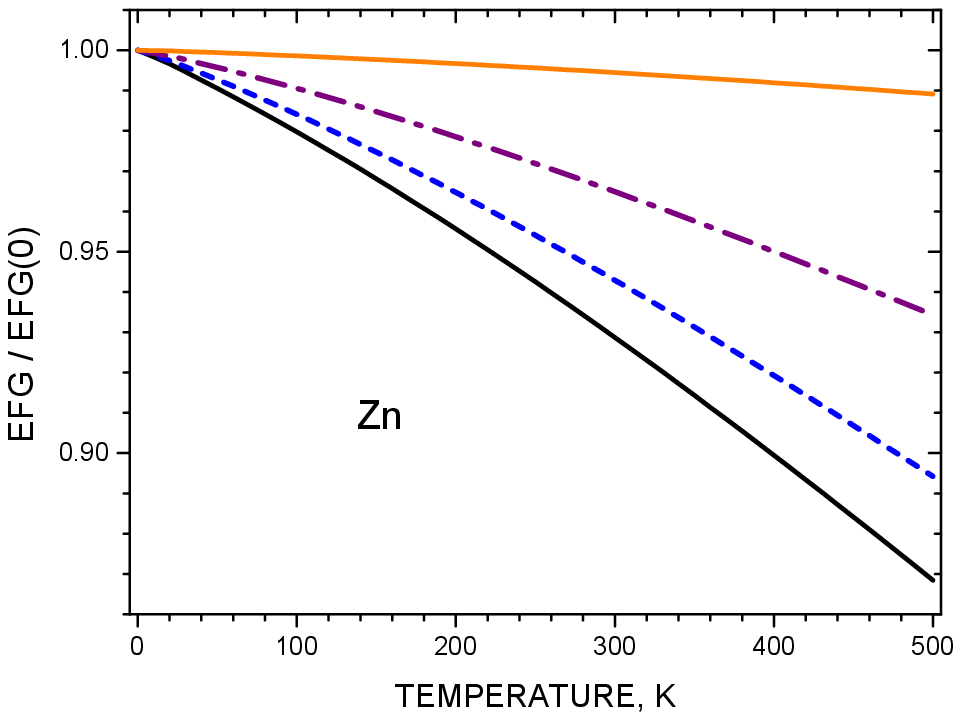}}
\vspace{2mm}
\resizebox{0.35\textwidth}{!}
{\includegraphics{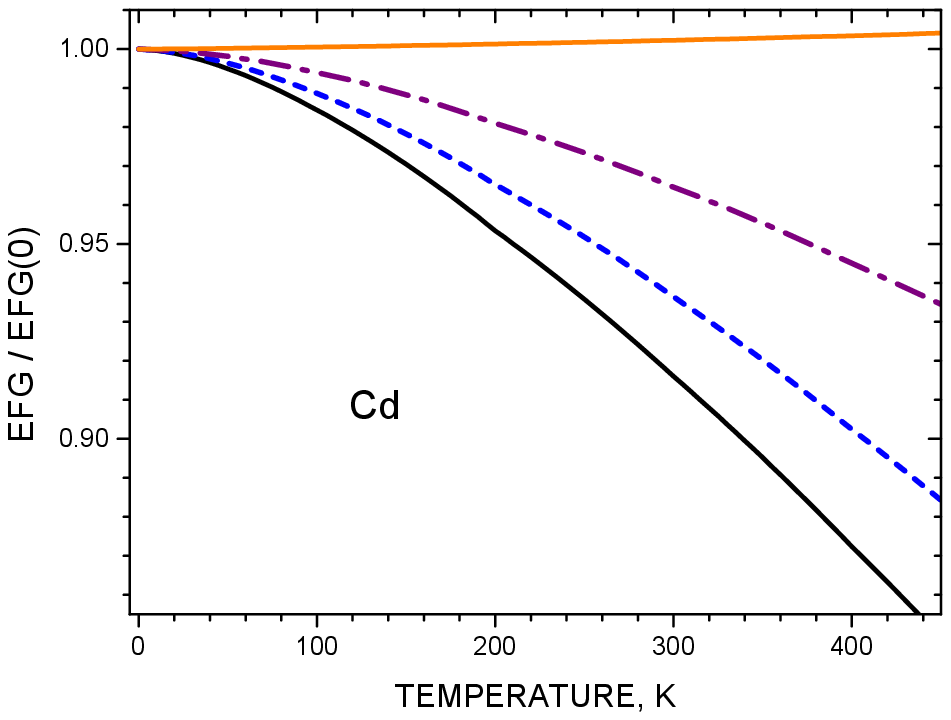}}
\caption{
Evolution of the $T$ dependence of $V_{zz}$ with a reduction of the mean square displacement ratio ${\cal R} = U_{33}/U_{11}$ while keeping the
averaged value of the displacements $U_{av}$, Eq.\ (\ref{zn1}), unchanged.
Black solid line corresponds to the initial value of ${\cal R}_0 = U_{33}/U_{11}$, the others - to reduced ${\cal R}$.
Blue dashed line: ${\cal R} = 1 + (2/3) \cdot ({\cal R}_0 - 1)$, purpled dot-dash line: ${\cal R} = 1 + (1/3)\cdot ({\cal R}_0 - 1)$,
orange solid line: ${\cal R}=1$ or $U_{33} = U_{11}$.
Top panel - zinc, bottom panel - cadmium. A weak increase of $V_{zz}$ with $T$ (orange solid line in Cd) is discussed in Sec.~\ref{incr}.
} \label{fig7}
\end{figure}

%
On the other hand, our calculations of the temperature evolution of EFG using experimental data for $U_{11}(T)$ and $U_{33}(T)$
demonstrate a good correspondence with the measured values of $V_{zz}$, Fig.\ \ref{fig6}.
We have also found that the $T$ dependence of $V_{zz}$ is extremely sensitive to the ratio ${\cal R} = U_{33}/U_{11}$.
To demonstrate it we have computed the evolution of $V_{zz}$ with $T$ with reduced values of ${\cal R}$,
Fig.\ \ref{fig7}(upper panel). It turns out that if ${\cal R} = 1$ [i.e. $U_{33}(T)=U_{11}(T)$] there is
virtually no temperature decrease of EFG, although we have kept the average value of mean-square displacements
\begin{eqnarray}
   U_{av} = (2U_{11} + U_{33})/3
  \label{zn1}
\end{eqnarray}
growing with $T$ according to experimental data \cite{Nus}.
It is worth mentioning that the experimental behavior of $U_{33}/U_{11}$ in Zn is highly nonlinear
in contrast to the simple linear increase of the lattice constant ratio $c/a$ \cite{Nus}.
On the other hand, in the harmonic approximation $U_{33}/U_{11}$ should be almost independent of $T$, at least for $T > T_D \sim 200$~K \cite{Ske}.
Therefore, in the $T$ dependence of $U_{33}/U_{11}$ and consequently of EFG
there is a substantial anharmonic contribution.
%
%
In particular, the anharmonic effect is responsible for the negative curvature of $V_{zz}$ in contrast to the approximate linear
dependence of the harmonic model, Fig.\ \ref{fig6}.

Notice, that even at zero temperature
the value of EFG is slightly reduced because of the zero point vibrations.
The calculated reduction is 0.6{\%}
with the mean square displacements from Ref.\ \cite{Nus} and 1{\%}
according to DFPT-QE.
The refined calculated absolute value of EFG is $V_{zz}=3.51\times 10^{21}$ V/m$^2$, 
the corrected for the zero point vibrations \cite{Nus} -- $V_{zz}=3.49\times 10^{21}$ V/m$^2$. 
It corresponds to the quadrupole frequency
10.53 MHz for $Q($Zn$)=0.125$~b \cite{Haa} and 12.64 MHz for $Q($Zn$)=0.15$~b, which compares well
with the experimental value of 12.34~MHz at 4.2~K \cite{Pot}.

\subsection{Cadmium}
\label{Cd}

\begin{figure}[!]
\resizebox{0.40\textwidth}{!}
{\includegraphics{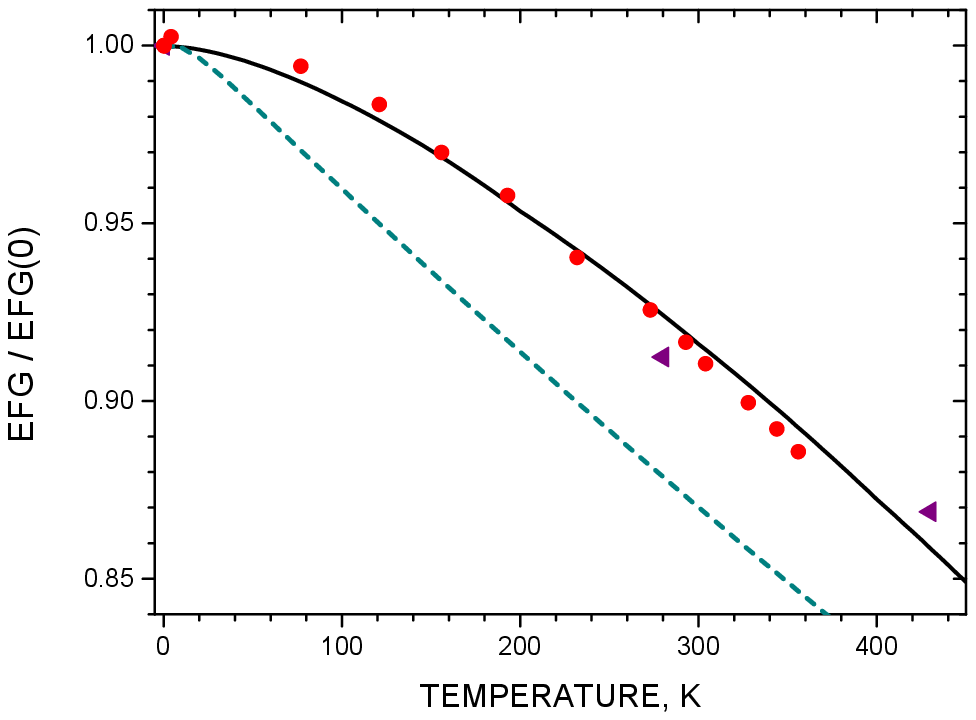}}
\vspace{2mm}
\resizebox{0.40\textwidth}{!}
{\includegraphics{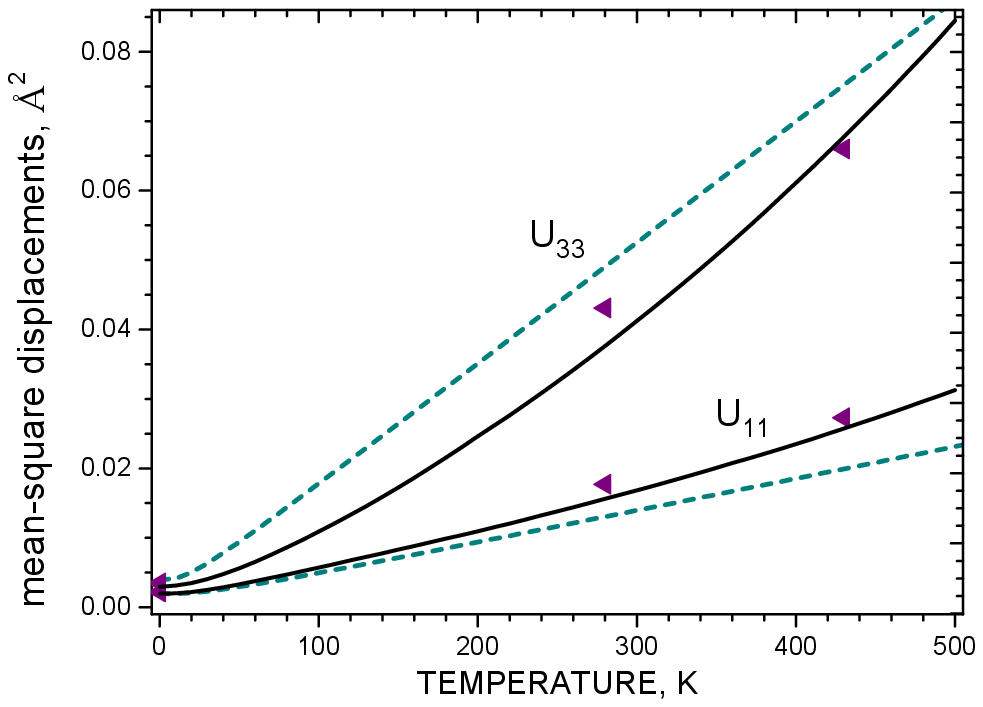}}

\vspace{2mm}
\caption{Temperature evolution of $V_{zz}$ in Cd (upper panel)
determined by the mean-square displacements $U_{11}$ and $U_{33}$ (lower panel).
Red circles stand for experimental measurements \cite{Chris},
black line -- for a present model calculation taking into account softening of the
lattice and an increase of $U_{33}/U_{11}$ with $T$ (see text for details),
dark cyan dash line -- DFPT-QE calculations \cite{QE1,QE2}.
Purple triangles are our calculations with mean-square displacements
of Table II of Ref.\ \cite{Toru} (the Phonon code \cite{phonon} within the harmonic approximation).
} \label{fig8}
\end{figure}

\begin{figure}[!]
\resizebox{0.35\textwidth}{!}
{\includegraphics{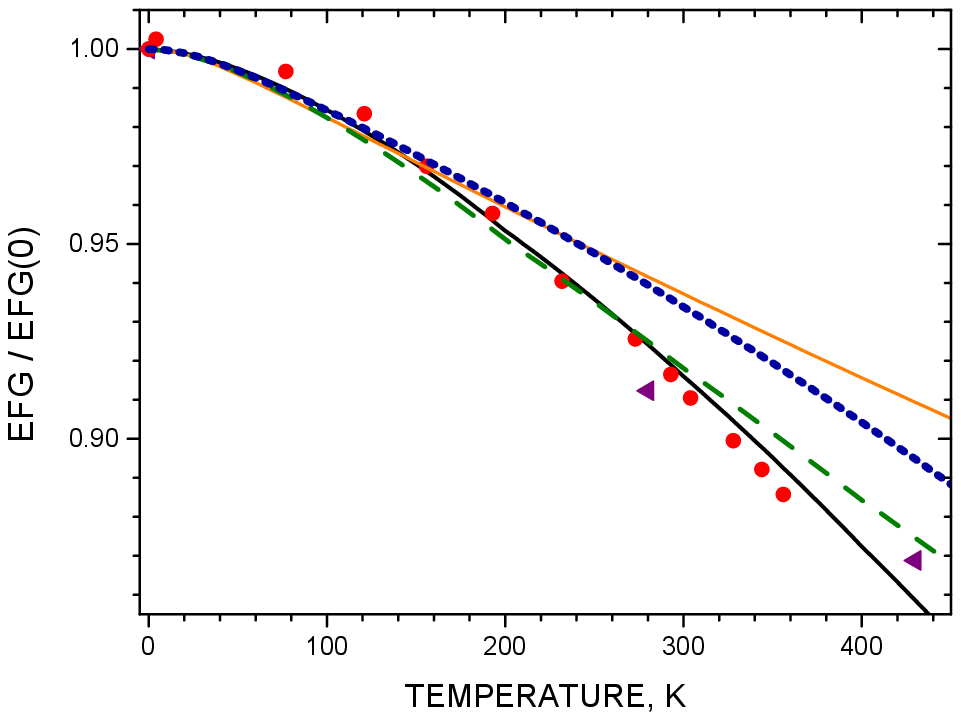}}
\vspace{2mm}
\resizebox{0.35\textwidth}{!}
{\includegraphics{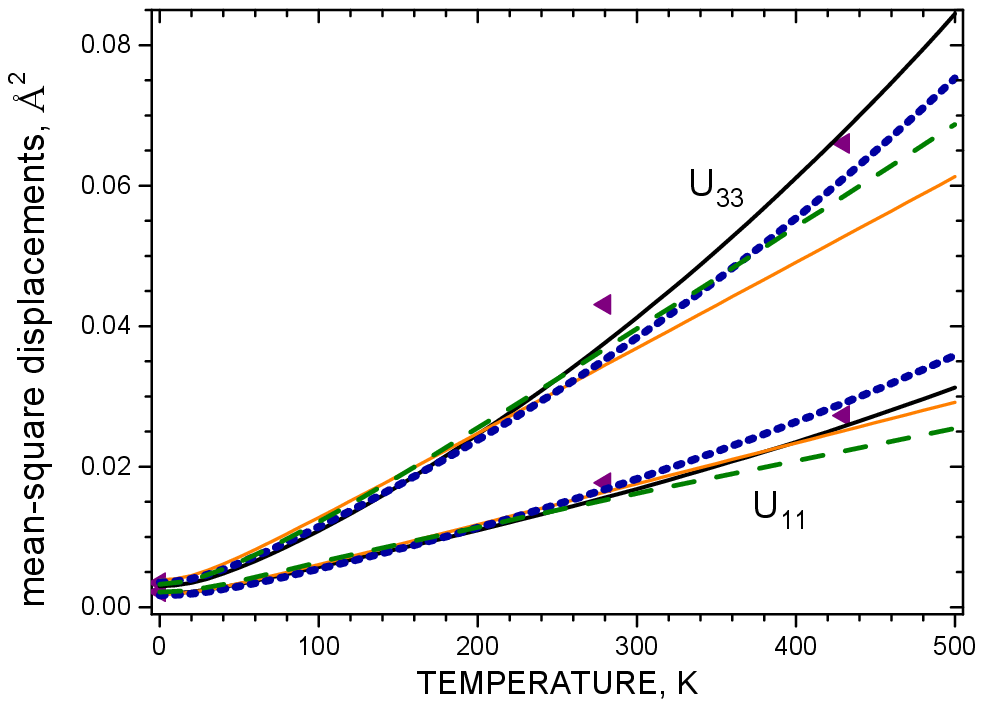}}

\vspace{2mm}
\caption{
Model mean-square displacements $U_{11}$ and $U_{33}$ in Cd (lower panel) and
corresponding temperature dependencies of $V_{zz}$ (upper panel).
Red circles stand for experimental measurements \cite{Chris},
black line -- for a model calculation taking into account the softening of the
lattice and the increase of $U_{33}/U_{11}$ with $T$.
Yellow solid line -- calculations without softening and the $U_{33}/U_{11}$ ratio fixed;
blue dotted line -- with the softening and the $U_{33}/U_{11}$ ratio fixed;
green dashed line -- without softening, the $U_{33}/U_{11}$ ratio increased,
see text for details.
Purple triangles are our calculations with mean-square displacements
of Table II of Ref.\ \cite{Toru}.
} \label{fig9}
\end{figure}

Results for cadmium are given in Figs.\ \ref{fig8} and \ref{fig9}.
%
%
As for the zinc calculation we observe that the mean-square displacements $U_{33}$ are
overestimated by the DFPT-QE calculations. In addition, DFPT-QE values of $U_{33}$ are underestimated
in comparison with other models.
Both effects lead to a fast decrease of $V_{zz}$ with temperature, although in the region of 260-360~K
the DFPT-QE harmonic treatment gives the correct slope for the $V_{zz}$ decrease.
As for zinc our calculations of the temperature evolution of $V_{zz}$ in Cd have turned out to be very sensitive to
values of mean square displacements $U_{11}$ and $U_{33}$, especially to ${\cal R} = U_{33}/U_{11}$.
It is worth noting that with the mean-square values obtained by Torumba et al \cite{Toru} with
the Phonon program \cite{phonon}, we have obtained reduced EFGs which compares well with
the experimental results of Ref.\ \cite{Chris} at 280 K and 430 K, Fig.\ \ref{fig8}.
However, our calculations also show that these values of $U_{11}$ and $U_{33}$
somewhat underestimate the reduction of $V_{zz}$ at 430 K and especially at 570 K (not shown in Fig.\ \ref{fig8}).
We believe that this is related to the softening of the Cd lattice \cite{Ske} and
increase of the ratio of ${\cal R}$ with temperature which lays beyond the harmonic
approximation.
In fact, as we know from experimentally deduced values of $U_{11}$ and $U_{33}$ \cite{Nus}
both effects are clearly present in the case of Zn lattice, Fig.\ \ref{fig4}.
Since at the moment there are no experimental data for the mean-square displacements
of cadmium, we have performed a number of model calculations, Fig.\ \ref{fig9}.
%
%
The averaged mean square displacement $U_{av} = (2U_{11} + U_{33})/3$ has been calculated within the Debye lattice model
with a temperature dependent Debye temperature $T_D$.
The lattice softening has been modelled by a linear decrease of $T_D$,
from 140 K (at $T=0$~K) to 115 K (at $T=500$~K), with the averaged value of $T_D^{av}=127.5$~K.
[These value is very close to $T_D=$131$\pm$7~K given in Ref.\ \cite{Kris} for Cd.]
In addition, we consider an anharmonic effect modelled by a linear change of ${\cal R}$ from 1.5 (at $T=0$~K) to 2.7 (at $T=500$~K)
with the averaged value of ${\cal R}=2.1$. Correspondingly, we have performed four model calculations shown in Fig.\ \ref{fig9}:
(1) with the lattice softening and the temperature increase of ${\cal R}$ (black curve in Fig.\ \ref{fig9}),
(2) without softening and with ${\cal R}$ fixed (yellow curve), (3) with the softening and ${\cal R}$ fixed (blue dotted line),
(4) without softening and with ${\cal R}$ increased (dashed green line).
The results clearly demonstrate that the best fit is achieved with the lattice softening and the temperature increase of ${\cal R}$ with $T$,
while other models deviate from the experimental data at elevated temperatures.
Although these arguments can not be considered as a solid proof, the important finding is that both unharmonic effects
improve the comparison with the experiment.

The calculated absolute value of EFG is $V_{zz}= 7.82\times 10^{21}$~V/m$^2$. The zero point vibration contribution to $V_{zz}$
amounts to 0.8{\%} according to DFPT-QE and 0.4{\%} in the model (1).
The corrected for the zero point vibrations
value [model (1)] is $V_{zz}= 7.79\times 10^{21}$~V/m$^2$, which
corresponds to $\nu_Q = 120.7$~MHz (adopting $Q($Cd$)=0.641$~b \cite{Haa}) or $\nu_Q = 143.1$~MHz (with $Q($Cd$)=0.76$~b \cite{Err}).
Extrapolated to $T=0$ quadrupole frequency in metallic cadmium is 136.9 MHz~\cite{Chris}.

\subsection{Mechanism of increase of EFG with $T$}
\label{incr}

The most important factor in the $T$ dependence of EFG is ${\cal R} = U_{33}/U_{11}$.
As discussed in Sec.\ \ref{Zn},~B and shown in Fig.\ \ref{fig7},
the ratio ${\cal R}$ has a pronounced influence on the shape of the $V_{zz}(T)$ curve.
A reduction of ${\cal R}$ in cadmium also has an immediate and sizeable effect on $V_{zz}(T)$, Fig.\ \ref{fig7}.
Interestingly, for ${\cal R} = 1$ [$U_{11} = U_{33}$] we observe in cadmium a rare effect of a weak increase of $V_{zz}$ with $T$
(orange plot in Fig.\ \ref{fig7}).
At first sight, this seems incompatible with the apparent reduction of $\langle \rho_Q \rangle$, Eq.~(\ref{e15}),
caused by $\exp(-W)$. Notice however, that in the sum on the right hand side of Eq.\ (\ref{e15}) individual contributions proportional to
$j_2(K R_{MT}) S_{Q}(\hat{K}) \langle \rho(\vec{K}) \rangle$
are of different signs. Two such principal groups of terms can be schematically written as
\begin{eqnarray}
 V_{zz}(T) \propto \langle \rho_Q \rangle = \sum [c_1\, {\cal W}_1\, \rho(\vec{K}_1) - c_2\, {\cal W}_2\,  \rho(\vec{K}_2) ] . \label{in1}
\end{eqnarray}
Here the $T$ dependent SRDWFs are incorporated in the weight factors ${\cal W}_i=\exp(-W(\vec{K}_i,\, T))$
and $c_i,\, \rho(\vec{K}_i) > 0$ ($i=1, 2$).
Both ${\cal W}_1$ and ${\cal W}_2$ reduce with $T$, but if
${\cal W}(\vec{K}_2,\, T)$ drop fast enough in comparison with ${\cal W}(\vec{K}_1,\, T)$, one can have a resulting increase
of the expression in the square brackets of (\ref{in1}) and of the whole sum.

\begin{figure}[!]
\resizebox{0.45\textwidth}{!}
{\includegraphics{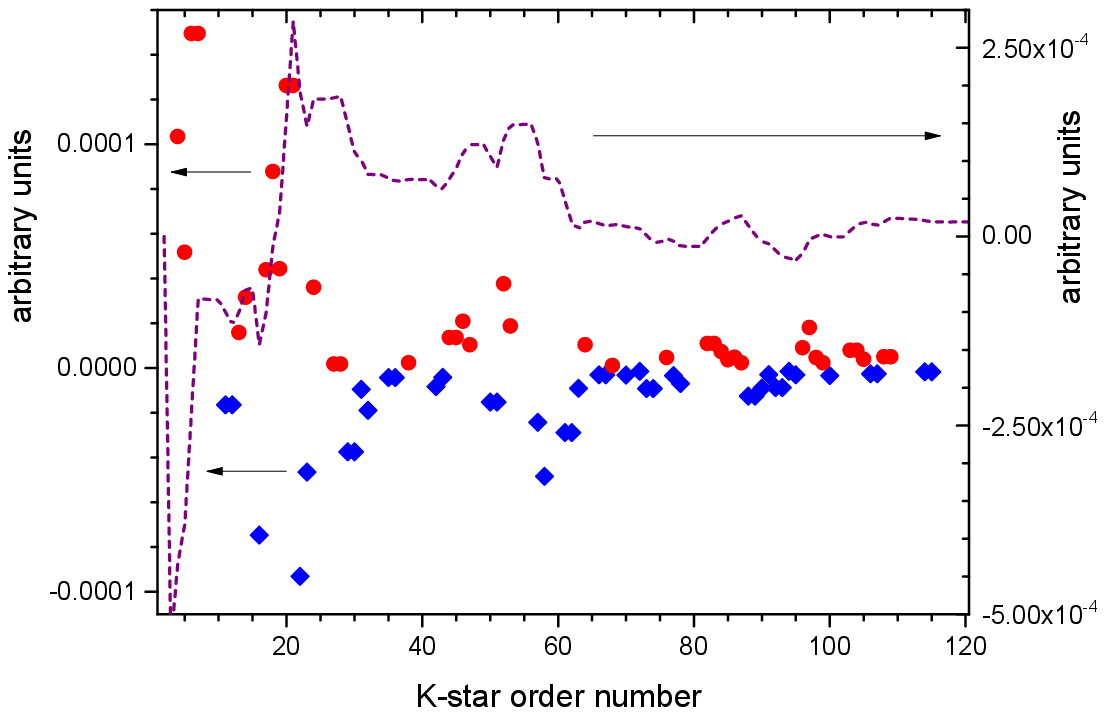}
}

\caption{
Individual $K-$star contributions
to electric field gradient at $T=300$~K [each term $\propto j_2(K R_{MT}) S_Q(\hat{K})
(\langle \rho(\vec{K}) \rangle_{T=300} - \langle \rho(\vec{K}) \rangle_{T=0})$, Eq.\ (\ref{e15})].
Blue diamonds - negative contributions, red circles - positive contributions,
dashed purpled curve - sum of all contributions up to the chosen $K-$star.
The final value of the sum (at $n \ge 120$) is positive, which leads to an increase
of $V_{zz}$ at 300~K.
} \label{fig10}
\end{figure}

Our detailed numerical analysis for the situation
is illustrated further in Fig.\ \ref{fig10}.
One can see that at $T=300$~K positive contributions to the gradient (red circles) although appreciably suppressed by
negative contributions (blue diamonds), finally prevail, which leads to a small positive net contribution to EFG
(the $n=120$ value of the dashed purpled curve). At $n \ge 120$ the positive sum value is virtually unchanged.
%
%

\section{Conclusions}
\label{sec:con}

%
\begin{figure}[!]
\resizebox{0.4\textwidth}{!}
{\includegraphics{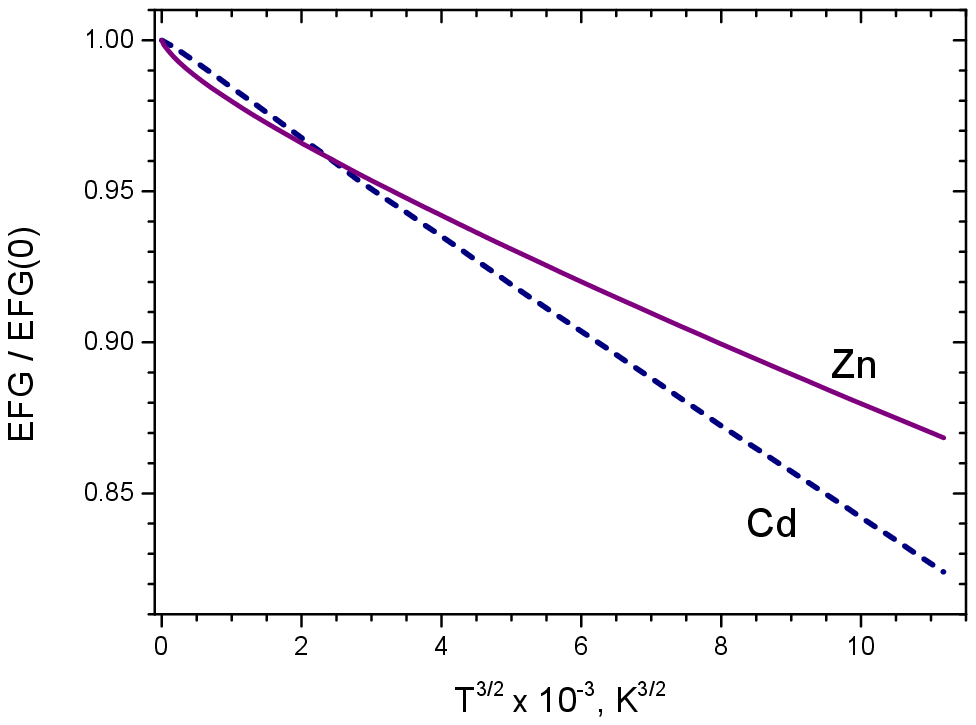}
}

\vspace{2mm}
\caption{
Plots of the $T$ dependence of electric field gradient, $V_{zz}(T)/V_{zz}(T=0)$ for Zn and Cd as functions of $T^{3/2}$.
} \label{fig11}
\end{figure}
%

We have presented a new method which can describe the temperature evolution of electric field gradient $V_{zz}$
in metals.
The consideration is based on the fact that the average value of the quadrupole component $\langle \rho_Q \rangle$ of electron density
on a sphere vibrating with the nucleus, Eq.\ (\ref{e15}), is changed with $T$.
The effective reduction of each Fourier component $\rho(\vec{K})$ of density, Eqs. (\ref{e11}), (\ref{e12}) and (\ref{e12'}),
is given by the multiplier $\exp(-W(\vec{K},T))$
which is the square root of the usual Debye-Waller factor.
The method relies on DFT calculations of electron properties and
the dependence of mean-square displacements $U_{11}$ and $U_{33}$ on temperature.
We have found that the form and pace of the temperature change of $V_{zz}$ is very sensitive to
the mean-square displacements $U_{11}(T)$, $U_{33}(T)$, and in particular to their ratio $U_{33}(T)/U_{11}(T)$.
The model is capable of reproducing both a decrease or increase of EFG with temperature.
The unusual mechanism of increase of $V_{zz}$ with $T$ is discussed in Sec.\ \ref{incr}.

We have applied our method to hexagonal close packed structures of pristine zinc and cadmium.
In the case of zinc where the mean-square displacements $U_{11}(T)$ and $U_{33}(T)$
are known experimentally from single crystal x-ray diffraction \cite{Nus},
we obtain a very good description of the temperature change of EFG, Fig.\ \ref{fig6}.
In the case of cadmium the behaviour of $V_{zz}(T)$, Fig.\ \ref{fig8}, can be reproduced
by assuming two anharmonic effects: the lattice softening (modelled by a decrease
of the Debye temperature $T_D$ with $T$), and an increase of the ratio $U_{33}/U_{11}$ with $T$.
It is worth noting that both anharmonic effects are also present in zinc \cite{Ske,Nus}.

In addition, we have performed calculations of the mean-square displacements in Zn and Cd
in the harmonic approximation by using the DF perturbation treatment of the QE package \cite{QE1,QE2}.
For both metals an approximately linear dependence in $T$ for $U_{11}(T)$ and $U_{33}(T)$ (at $T > 40$~K)
and consequently for the calculated $V_{zz}$ curves has been found.
In the case of cadmium, the decrease of EFG has been exaggerated which can be accounted for by
an uncertainty related to its core pseudopotential \cite{sssp1,sssp2}.
Using the mean-square displacements calculated at $T = 280$ and 430~K
by Torumba et al. \cite{Toru} with the package Phonon \cite{phonon},
we have obtained reduced values of EFG which lie not far from the experimental data.

In our studies we have not found an intrinsic mechanism for the $T^{3/2}$ dependence of EFG, Eq.\ (\ref{i2}).
Nevertheless, the $T^{3/2}$ plots, Fig.\ \ref{fig11}, indicate that an approximate $T^{3/2}$ law
for Zn and Cd holds. For Cd the dependence is almost perfect, for Zn it deviates
from $T^{3/2}$ at low temperatures which can be partially explained by the fact that the experimental data
for mean-square displacements \cite{Nus} are available only for $T > 40$~K.
In any case, even experimental data for EFG of Zn and Cd deviate from the $T^{3/2}$ law at low temperatures \cite{Verma}.
We therefore conclude that the mechanism of the temperature dependence of $V_{zz}$ in Zn and Cd
is complex, with a substantial contribution from anharmonic effects.
An approximate correspondence with the $T^{3/2}$ dependence is probably
due to the $T$ behavior of SRDW factors, $\exp(-W)$ \cite{Jen}.

Finally, we mention that even at zero temperature the measured EFGs are smaller than $V_{zz}$
calculated by {\it ab initio} methods.
The zero temperature reduction however is small: 0.6{\%} in Zn
and 0.4{\%} in Cd.

\acknowledgements
The work is supported by a Polish representative in the Joint Institute for Nuclear Research.
The {\it ab initio} investigation  of phonon frequencies was supported by the Russian Science Foundation (grant 18-12-00438). Calculations  were carried out using computing resources of the federal collective usage center `Complex for Simulation and Data Processing for Mega-science Facilities' at NRC `Kurchatov Institute' (http://ckp.nrcki.ru/), supercomputers at Joint Supercomputer Center of Russian Academy of Sciences (http://www.jscc.ru)  and ``Uran'' supercomputer of IMM UB RAS (http://parallel.uran.ru).


\appendix

\section{Quadrupolar potential and the tensor of EFG}
\label{append_A}

The asymptotic behavior of the potential close to the nucleus, Eq.\ (\ref{p2}),
can be understood from the two center expansion of the Coulomb potential in double multipolar series,
\begin{eqnarray}
 & & \frac{1}{|\vec{R}(\vec{n})-\vec{R}'(\vec{n}')|} \nonumber \\
 & & = \sum_{\Lambda \Lambda'} v_{\Lambda \Lambda'}(\vec{n},\vec{n}';\,r,r')\,
    S_{\Lambda}(\hat{r}(\vec{n}))\, S_{\Lambda'}(\hat{r}'(\vec{n}')) ,
 \label{p3a}
\end{eqnarray}
where the interaction strength $v_{\Lambda \Lambda'}$ is given by \cite{Hei}
\begin{eqnarray}
 v_{\Lambda \Lambda'}(\vec{n},\vec{n}';\,r,r')
 \sim \frac{(r)^l (r')^{l'}}{|\vec{X}(\vec{n})-\vec{X}(\vec{n}')|^{l+l'+1}},
 \label{p3b}
\end{eqnarray}
Here $\vec{R}(\vec{n})=\vec{X}(\vec{n}) + \vec{r}(\vec{n})$ is the radius vector close to the crystal site $\vec{n}$.
For $r \ll 1$ we arrive at Eq.\ (\ref{p2}) which holds both for
the contributions from the electron density around the site $\vec{n}$ and the contributions from the densities on the other sites $\vec{n'} \neq \vec{n}$.

As follows from Eqs.\ (\ref{p1}), (\ref{p2}),
the quadrupolar component of the potential for the hexagonal lattice
can be written as
\begin{eqnarray}
 V_{Q}(r,\theta,\phi) &=& v_{Q}\, r^2\, Y_{l=2}^{m=0}(\theta,\phi) \nonumber \\
                      &=& v_{Q}\,\frac{1}{4} \sqrt{\frac{5}{\pi}} (3z^2 - r^2) .
 \label{p4}
\end{eqnarray}
[We recall that here $v_{Q} = v_{(2,1)}$ is a constant.]
From this relation we obtain that the tensor of EFG is diagonal in the Cartesian system of axis and
\begin{eqnarray}
 & &V_{zz} = \frac{ \partial^2 V_{Q}}{\partial z^2} = \sqrt{\frac{5}{\pi}} \, v_{Q} , \label{p5a} \\
 & &V_{xx} = \frac{ \partial^2 V_{Q}}{\partial x^2} = \frac{ \partial^2 V_{Q}}{\partial y^2} = -\frac{1}{2} \sqrt{\frac{5}{\pi}} \, v_{Q} . \label{p5b}
\end{eqnarray}


\end{document}